\documentclass[reprint,
 amsmath,amssymb,
 aps,
]{revtex4-2}

\usepackage{graphicx}
\usepackage{dcolumn}
\usepackage{bm}
\usepackage{xcolor}
\usepackage{mathptmx}
\usepackage{graphicx}
\usepackage{dcolumn}
 
\usepackage{bm}
\usepackage{amssymb}
\usepackage{comment}
\usepackage{hyperref}

\begin{document}

\title{Bending space-time wave packets}

\author{Layton A. Hall$^{1,2}$}

\author{Ayman F. Abouraddy$^{1,*}$}

\affiliation{$^{1}$CREOL, The College of Optics \& Photonics, University of Central~Florida, Orlando, FL 32816, USA}
\affiliation{$^{2}$Current address: Materials Physics and Applications - Quantum Division, Los Alamos National Laboratory, Los Alamos, NM 87545, USA}
\affiliation{$^*$Corresponding author: raddy@creol.ucf.edu}

\begin{abstract}
Optical beams with certain asymmetric profiles, such as the Airy beam, can depart from rectilinear propagation and instead travel along curved (typically parabolic) trajectories. Here we show that sculpting the spatiotemporal spectrum of optical pulses yields self-accelerating beams that have symmetric profiles, remain diffraction-free, and travel along power-law curves with propagation distance having arbitrary positive exponent (integer or fractional). We build upon propagation-invariant space-time wave packets (STWPs), in which each spatial frequency is associated with a single wavelength. A linear tilt in the propagation path of an STWP is produced by a corresponding tilt in the spectral domain. A curved trajectory is then produced through locally changing the tilt direction along the propagation axis, which requires associating a prescribed finite-bandwidth spatial spectrum to each wavelength. Using this approach, we realize symmetric STWPs traveling along curved trajectories that follow linear, quadratic, cubic, or even square-root power laws with an acceleration rate that is independent of the beam spatial scale. These novel bending STWPs open new avenues for realizing target-avoidance with electromagnetic waves.
\end{abstract}


\maketitle

\begin{figure}[t!]
\centering
\includegraphics[width=8.6cm]{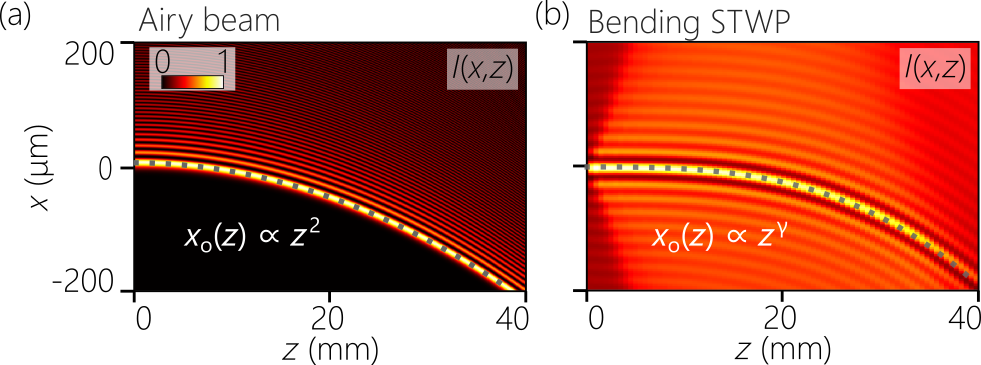}
\caption{(a) Intensity $I(x,z)$ of an Airy beam travels along a parabolic trajectory (dotted black curve); $x_{\mathrm{o}}(z)\propto z^{2}$. (b) The time-averaged intensity $I(x,z)$ of a bending STWP with a \textit{symmetric} profile travels along a curved trajectory, $x_{\mathrm{o}}(z)\propto z^{\gamma}$, where $\gamma>0$ (here $\gamma=3$, dotted black curve).}
\label{Fig:Concept}
\end{figure}

The rectilinear propagation of light rays is the foundation of geometric optics. As the transverse scale is reduced, wave optics introduces diffractive spatial spreading around the straight-line geometric trajectory of an underlying ray. An exception for this general rule is so-called `self-accelerating' optical beams, such as Airy beams \cite{Siviloglou07OL} whose transverse spatial profile conforms to an (asymmetric) Airy function \cite{Siviloglou07PRL}, in addition to other beams that display similar behavior \cite{Bandres08OL,Davis09OE,Efremidis19Optica}; see Fig.~\ref{Fig:Concept}(a). The transverse profile is translated laterally with axial propagation along a parabolic trajectory while remaining self-similar. These self-accelerating beams have been used to produce curved channels in plasmas \cite{Polynkin09Science}, to guide curved electric arcs around objects \cite{Clerici15SciAdv}, and in microscopy \cite{Vettenburg2014NM,Subedi2021OE}, among other possibilities \cite{Kaminer15NPhys,Efremidis19Optica}. This unusual propagation behavior immediately suggests applications in laser target-avoidance by bending the laser beam around an object, which has not been confirmed yet.

The transverse acceleration of an Airy beam is tied to its asymmetric profile, which determines the orientation of its curved trajectory. Moreover, the rate of acceleration is determined by the transverse spatial scale of the Airy beam. These are general features of the self-accelerating monochromatic beams realized to date. A different strategy that has recently emerged for sculpting optical fields introduces spatiotemporal coupling into a pulsed beam to modify the field behavior and produces exotic propagation characteristics. For example, space-time wave packets (STWPs), in which a tight association is introduced between the spatial and temporal frequencies underpinning the pulsed beam \cite{Kondakci17NP,Yessenov22NC,Yessenov22AOP}, display propagation invariance with controllable group velocity in linear media \cite{Wong17ACSP2,Efremidis17OL,Porras01OL,Kondakci19NC}, anomalous refraction \cite{Bhaduri20NP}, self-healing \cite{Kondakci18OL}, and dispersion cancellation \cite{He22LPR,Hall23LPR}. Loosening the tight association between the spatial and temporal frequencies can introduce axial variation in a parameter of the STWP, including axial acceleration \cite{Clerici08OE,Yessenov20PRLAccel,Li20SR,Hall22OLAccel}, and axial spectral encoding \cite{Motz21PRA,Hall25OL}. Nevertheless, these fields travel rectilinearly. A recent theoretical study suggests a modification to propagation-invariant STWPs that curves the propagation trajectory \cite{Liang23OL}.

Here we demonstrate experimentally that judicious spatiotemporal structuring of STWPs localized along one transverse dimension (similarly to Airy beams) can produce curved trajectories while maintaining a spatially symmetric transverse profile [Fig.~\ref{Fig:Concept}(b)], which we call `bending STWPs'. We present an algorithmic procedure to design the 2D phase distribution used to modulate the spatiotemporal spectrum of a generic pulsed beam and produce a bending STWP. The resulting curved propagation trajectory follows a power law in the axial distance with a positive exponent. As examples, we experimentally demonstrate linear, quadratic, cubic, and even square-root exponents of the power law. Moreover, these power laws are all associated with the \textit{same} transverse profile shape and scale. These results suggest spatiotemporal structuring of light as a methodology for bending the trajectory of a beam.

\begin{figure}[t!]
\centering
\includegraphics[width=8.6cm]{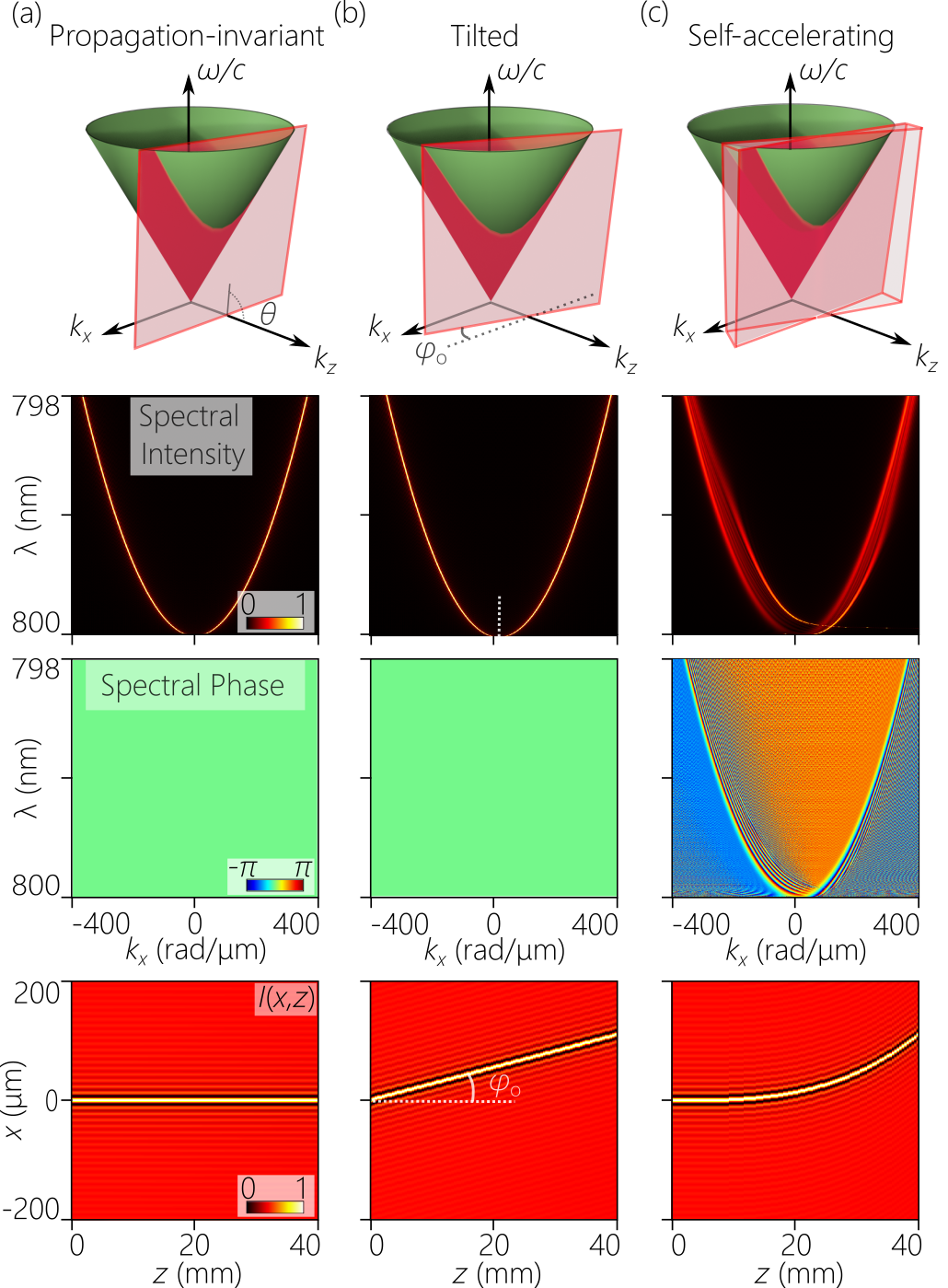}
\caption{(a) A propagation-invariant STWP. (b) A tilted STWP with $\varphi_{\mathrm{o}}=0.15^{\circ}$. (c) A bending STWP with $x_{\mathrm{o}}(z)=x_{1}(\tfrac{z}{z_{1}})^{3}$, where $x_{1}=100$~$\mu$m and $z_{1}=40$~mm. The first row is the STWP spectral support on the light-cone; the second and third rows are the spectral intensity and phase projected onto the $(k_{x},\lambda)$-plane, respectively; and the fourth row the time-averaged intensity $I(x,z)$. Throughout, $\theta\!=\!60^{\circ}$ and $\Delta\lambda\!=\!2$~nm.}
\label{Fig:SpatiotemporalSpectra}
\end{figure}

\textbf{Propagation-invariant STWPs.} We consider wave packets propagating along the $z$-axis in which only one transverse coordinate $x$ is relevant (the field is uniform along $y$). Propagation-invariant STWPs [Fig.~\ref{Fig:SpatiotemporalSpectra}(a)] inculcate the following restriction on the axial wave numbers: $k_{z}-k_{\mathrm{o}}=\Omega/\widetilde{v}$ \cite{Yessenov22AOP}, where $\omega$ is the temporal frequency, $\omega_{\mathrm{o}}$ is a fixed temporal frequency, $\Omega=\omega-\omega_{\mathrm{o}}$, $k_{\mathrm{o}}=\omega_{\mathrm{o}}/c$, $c$ is the speed of light in vacuum, and $\widetilde{v}$ is the STWP group velocity. Therefore, the spectral support for such an STWP is the intersection of the free-space light-cone $k_{x}^{2}+k_{z}^{2}=(\tfrac{\omega}{c})^{2}$ with a plane that is parallel to the $k_{x}$-axis and makes an angle $\theta$ (the spectral tilt angle) with the $k_{z}$-axis, corresponding to a parabolic spatiotemporal spectrum $\tfrac{\Omega}{\omega_{\mathrm{o}}}=\tfrac{k_{x}^{2}}{2k_{\mathrm{o}}^{2}(1-\cot\theta)}$ in the paraxial regime. The STWP field $E(x,z;t)=e^{i(k_{\mathrm{o}}z-\omega_{\mathrm{o}}t)}\psi(x,z;t)$ has an envelope that travels rigidly in free space at a group velocity $\widetilde{v}=c\tan\theta$, $\psi(x,z;t)=\psi(x,0;t-z/\widetilde{v})$. The $(k_{x},\tfrac{\omega}{c})$-projection is a parabola and the $(k_{z},\tfrac{\omega}{c})$-projection is a straight line. The time-averaged intensity $I(x,z)=\int dt|E(x,z;t)|^{2}=I(x,0)$ is diffraction-free along the $z$-axis [Fig.~\ref{Fig:SpatiotemporalSpectra}(a)].  We denote the coordinate of the profile center $x_{\mathrm{o}}(z)$, with $x_{\mathrm{o}}(0)=0$. For a propagation-invariant STWP we have $x_{\mathrm{o}}(z)=0$.

\textbf{Tilted STWPs.} The propagation path of a propagation-invariant STWP can nevertheless be tilted away from the $z$-axis by an angle $\varphi_{\mathrm{o}}$, such that $z'=z\cos\varphi_{\mathrm{o}}-x\sin\varphi_{\mathrm{o}}$ and $x'=z\sin\varphi_{\mathrm{o}}+x\cos\varphi_{\mathrm{o}}$, where $(x',z')$ is the new coordinate system in which the STWP is tilted [Fig.~\ref{Fig:SpatiotemporalSpectra}(b)]. This coordinate transformation in physical space is associated with a rotation in Fourier space: $k_{z}'=k_{z}\cos\varphi_{\mathrm{o}}-k_{x}\sin\varphi_{\mathrm{o}}$ and $k_{x}'=k_{z}\sin\varphi_{\mathrm{o}}+k_{x}\cos\varphi_{\mathrm{o}}$, which requires the plane containing the STWP spectral support to be also rotated around the $\tfrac{\omega}{c}$-axis by $\varphi_{\mathrm{o}}$ [Fig.~\ref{Fig:SpatiotemporalSpectra}(b)]. The spectral support is still a parabola, but its $(k_{x},\tfrac{\omega}{c})$-projection is no longer symmetric around $k_{x}=0$, and its $(k_{z},\tfrac{\omega}{c})$-projection is no longer a straight line. The transverse profile of the time-averaged intensity $I(x,z)$ is diffraction-free but the propagation trajectory is tilted with the $z$-axis $x_{\mathrm{o}}(z)=z\tan\varphi_{\mathrm{o}}$ [Fig.~\ref{Fig:SpatiotemporalSpectra}(b)].

\textbf{Bending STWPs.} An example of a bending STWP is shown in Fig.~\ref{Fig:SpatiotemporalSpectra}(c), whose center travels along a trajectory following a cubic law, while maintaining a symmetric profile almost identical to that of the STWPs in Fig.~\ref{Fig:SpatiotemporalSpectra}(a,b). The spatial profile of the time-averaged intensity $I(x,z)$ therefore does not indicate the orientation of the curved trajectory. We pursue here trajectories of the form $x_{\mathrm{o}}(z)=x_{1}(\tfrac{z}{z_{1}})^{\gamma}$, where $\gamma\geq0$ (whether integer, fractional, or real), and $x_{1}$ and $z_{1}$ are transverse and axial scaling factors, respectively. Heuristically, this bending trajectory requires a continuous axially varying tilt angle $\varphi(z)$ for the STWP spectral support around the $\tfrac{\omega}{c}$-axis. Consequently, the spectral support on the light-cone surface is no longer a curve, but rather a 2D domain formed from the continuous displacement of the parabola associated with a propagation-invariant STWP. Rather than give a formula for this spectral support, we instead pursue an algorithmic procedure for synthesizing bending STWPs, which requires understanding the experimental arrangement.

\begin{figure}[t!]
\centering
\includegraphics[width=8.6cm]{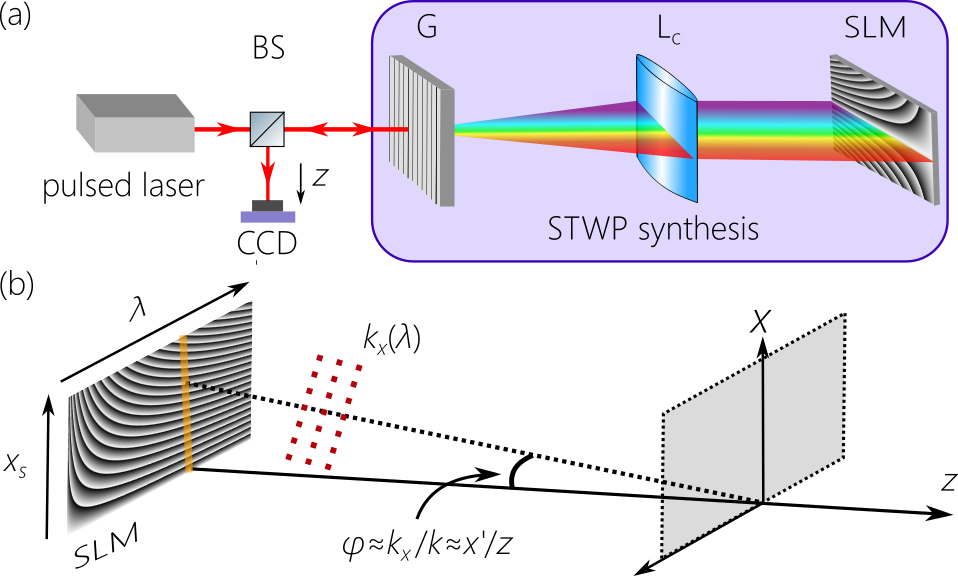}
\caption{(a) Schematic of the setup for synthesizing STWPs. G: Diffraction grating; L$_{\mathrm{c}}$: cylindrical lens; SLM: spatial light modulator. (b) Conceptual diagram for the axial mapping of spatial frequencies encoded at the SLM, which underpins the transverse self-acceleration of bending STWPs.}
\label{Fig:SetupAndSLM}
\end{figure}

\begin{figure*}[t!]
\centering
\includegraphics[width=16.0cm]{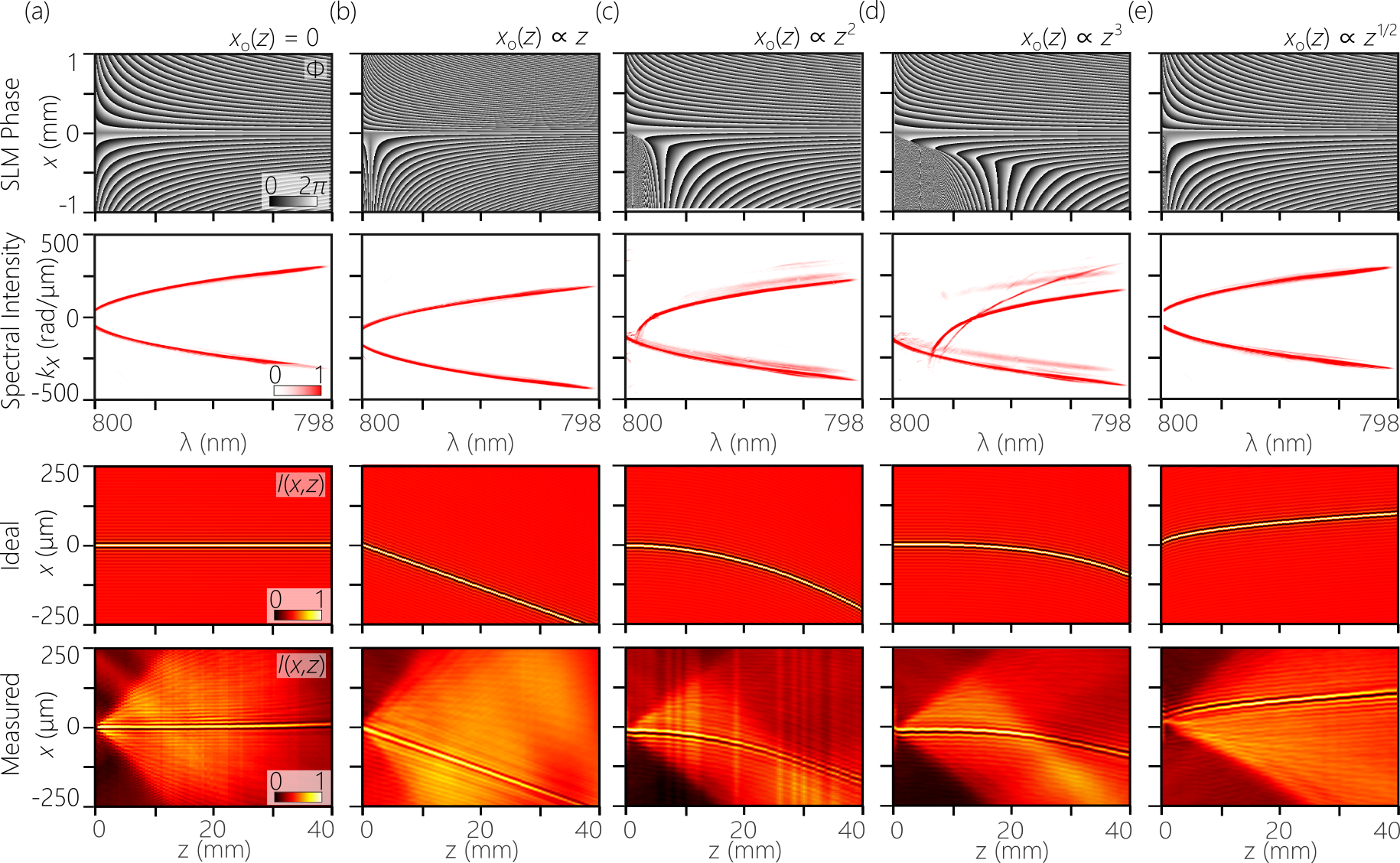}
\caption{The first row displays the SLM phase distribution $\Phi(\lambda,x)$; the second row the measured spatiotemporal spectral intensity; the third row the target time-averaged intensity profile $I(x,z)$; and the fourth row the measured $I(x,z)$. Throughout we have $\theta\!=\!60^{\circ}$, $\Delta\lambda\!=\!2$~nm, $\lambda_{\mathrm{o}}\!=\!800$~nm, and $z_{1}=40$~mm. (a) A propagation-invariant STWP; (b) a tilted STWP with $\varphi_{\mathrm{o}}=0.35^{\circ}$; (c) a bending STWP with $x_{\mathrm{o}}(z)=x_{1}(\tfrac{z}{z_{1}})^{2}$ and $x_{1}=200$~$\mu$m; (d) a bending STWP with $x_{\mathrm{o}}(z)=x_{1}(\tfrac{z}{z_{1}})^{3}$ and $x_{1}=150$~$\mu$m; and (e) a bending STWP with $x_{\mathrm{o}}(z)=x_{1}(\tfrac{z}{z_{1}})^{0.5}$ and $x_{1}=100$~$\mu$m.}
\label{Fig:labexp}
\end{figure*}

\textbf{Experimental setup.} We make use of the universal angular dispersion (AD) synthesizer in Refs.~\cite{Hall24JOSAA,Romer25JO}; see Fig.~\ref{Fig:SetupAndSLM}(a). The source is a mode-locked Ti:sapphire laser (Tsunami; Spectra-Physics) with a pulse width $\sim100$~fs at $\lambda_{\mathrm{o}}\sim800$~nm. The beam diameter is expanded to 25~mm and directed to a diffraction grating (Newport, 10HG1200-800-1; 1200~lines/mm, of area 25$\times$25~mm$^2$) to spatially resolve the pulse spectrum. The -1 diffraction order is collimated with a cylindrical lens (focal length $f=500$~mm), and a spatial light modulator (SLM, Hamamatsu X10468-02) placed at its focal plane imparts a 2D phase distribution $\Phi(\lambda,x_{\mathrm{s}})$ to the wave front, which associates each wavelength $\lambda$ with a spatial phase; here $x_{\mathrm{s}}$ is the SLM dimension orthogonal to the spectral axis. The retro-reflected field is recombined at the grating, thereby producing the STWP.

\textbf{Algorithm for designing the spatiotemporal spectral phase.} For propagation-invariant STWPs, the SLM phase distribution associated with $\omega$ is linear in $x_{\mathrm{s}}$, $\Phi(\omega,x_{\mathrm{s}})=k_{x}(\omega)x$, where $k_{x}(\omega)=\pm k_{\mathrm{o}}\sqrt{2\tfrac{\Omega}{\omega_{\mathrm{o}}}(1-\cot\theta)}$. This phase is displayed along the SLM column corresponding to $\omega$, and the positive and negative values of $k_{x}(\omega)$ occupy the lower and upper halves of the SLM, respectively. To rotate the spectral plane an angle $\varphi$ around the $\tfrac{\omega}{c}$-axis, we calculate a new spatial frequency $k_{x}'(\omega)=k_{z}(\omega)\sin\varphi+k_{x}(\omega)\cos\varphi$. To make this rotation $z$-dependent, we note that different portions along $x_{\mathrm{s}}$ on the SLM map to axial positions through $\tfrac{x_{\mathrm{s}}}{z}\sim\tfrac{k_{x}'}{k_{z}'}$ \cite{Motz21PRA,Hall25OL}; see Fig.~\ref{Fig:SetupAndSLM}(b).

The algorithmic procedure for synthesizing bending STWPs is as follows: (1) Select the parameters of the underlying STWP: $\theta$, $\lambda_{\mathrm{o}}$, and the bandwidth $\Delta\lambda$, which determine the spatiotemporal spectrum $k_{x}(\omega)$, and the beam profile shape and scale. (2) Select the curved trajectory $x_{\mathrm{o}}(z)=x_{1}(\tfrac{z}{z_{1}})^{\gamma}$, and determine the updated spatiotemporal spectrum $k_{x}'(\omega,z)$, which is now $z$-dependent. (3) The position $x_{\mathrm{s}}$ on the SLM at which $k_{x}'(\omega,z)$ is displayed is determined from $\tfrac{x_{\mathrm{s}}}{z}\sim\tfrac{k_{x}'(\omega,z)}{k_{z}'}$, which allows us to replace $z$ with $x_{\mathrm{s}}$ in $k_{x}'$, and the SLM phase now takes the form $\Phi(\omega,x_{\mathrm{s}})=k_{x}'(\omega,x_{\mathrm{s}})x_{\mathrm{s}}$. Repeating these steps for each $\omega$ (each column on the SLM) allows us to design the full SLM phase.

\textbf{Measurement results.} We start with producing a propagation-invariant STWP with $\theta=60^{\circ}$, $\lambda_{\mathrm{o}}\approx800$~nm, and bandwidth $\Delta\lambda\approx2$~nm [Fig.~\ref{Fig:labexp}(a)]. We plot the SLM phase $\Phi(\lambda,x_{\mathrm{s}})=k_{x}(\lambda)x_{\mathrm{s}}$, where $k_{x}(\lambda)$ is independent of $x_{\mathrm{s}}$. The measured spatiotemporal spectrum in the $(k_{x},\lambda)$-plane is a parabola that is symmetric around $k_{x}=0$, and the time-averaged intensity shows a symmetric beam traveling invariantly along a straight line, so that $x_{\mathrm{o}}(z)=0$. The central feature has a width $\Delta x\approx 10$~$\mu$m that travels without diffraction for $\approx 60$~mm.

An STWP tilted an angle $\varphi_{\mathrm{o}}=0.35^{\circ}$ with the $z$-axis is depicted in Fig.~\ref{Fig:labexp}(b). Retaining the parameters of the propagation-invariant STWP, we rotate the entire spectral support on the light-cone by $\varphi_{\mathrm{o}}$ around the $\tfrac{\omega}{c}$-axis. This rotation displaces the $(k_{x},\lambda)$ projection, which is no longer centered around $k_{x}=0$, and the direction of displacement indicates the direction of the beam tilt [Fig.~\ref{Fig:labexp}(b), second row]. The beam retains its symmetric profile, but its peak travels along a straight trajectory tilted to the $z$-axis, $x_{\mathrm{o}}(z)=(\tan\varphi_{\mathrm{o}})z$.

In Fig.~\ref{Fig:labexp}(c) we present a bending STWP following a parabolic trajectory $x_{\mathrm{o}}(z)=x_{1}(\tfrac{z}{z_{1}})^{2}$, with $x_{1}=200$~$\mu$m and $z_{1}=40$~mm. A significant change occurs in the SLM phase $\Phi(\lambda,x_{\mathrm{s}})$ because the phase associated with each $\lambda$ is no longer linear in $x_{\mathrm{s}}$. Moreover, the mirror symmetry between the upper and lower SLM halves is broken, as in the previous case of a tilted STWP, which signifies that the STWP no longer travels along the $z$-axis. The $(k_{x},\lambda)$-projection is displaced from $k_{x}=0$ and is no longer a curve. Rather, it is broadened, so that the relationship between $k_{x}$ and $\lambda$ is not one-to-one. The transverse beam profile is still symmetric, but its peak now travels along a parabolic trajectory in a similar manner to an Airy beam.

Uniquely, any power law, besides parabolic, can be realized. We present in Fig.~\ref{Fig:labexp}(d,e) two more examples of bending STWPs. A cubic trajectory of the form $x_{1}(z)=x_{1}(\tfrac{z}{z_{1}})^{3}$, with $x_{1}=200$~$\mu$m and $z_{1}=40$~mm, is shown in Fig.~\ref{Fig:labexp}(d). The phase pattern is drastically different from the STWP in Fig.~\ref{Fig:labexp}(a). Furthermore, the self-acceleration is not restricted to an integer-exponent power law. We plot in Fig.~\ref{Fig:labexp}(e) the results for a trajectory of the form $x_{1}(z)=x_{1}(\tfrac{z}{z_{1}})^{1/2}$, where $x_{1}=100$~$\mu$m and $z_{1}=40$~mm. This is -- to the best of our knowledge -- the first curved-trajectory self-accelerating optical beam that depends on the square-root of the propagation-distance.

\textbf{Discussion.} Airy-like beams have an asymmetric profile that fuels their self-acceleration and indicates the orientation of the beam curvature, and the spatial scale determines the acceleration rate. The beams we have produced here via spatiotemporal structuring circumvent these restrictions: the bending STWPs have a symmetric profile, the orientation of the beam curvature is not determined by the initial beam profile, and the transverse scale does not determine the acceleration rate. In contrast to Airy beams whose spatial profile is predetermined, the profile of bending STWPs can be modified, without affecting the self-acceleration, via spectral phase modulation \cite{Kondakci18PRL,Wong21OE}. Although we have made use of a pulsed laser, incoherent light can also be used to produce bending STWPs \cite{Yessenov19Optica,Yessenov19OL}.

Combining control over the transverse \textit{and} axial acceleration \cite{Yessenov20PRLAccel,Hall22OLAccel} of STWPs opens the path towards studying spatiotemporal self-acceleration \cite{Liang23OL}. Recently, it has been appreciated that STWPs can be viewed as a consequence of implementing a Lorentz boost on a generic focused beam \cite{Saari04PRE,Longhi04OE,Yessenov23PRA,Ramsey23PRA}, so that STWPs constitute table-top studies of special relativistic effects \cite{Hall23NP,Yessenov24PRA}. We envision that self-accelerating STWPs may enable experimental emulation of gravitational effects using optical fields.

Several avenues of future research can be envisioned based on the results presented here, including: (1) extending the propagation distance to the kilometer scale \cite{Hall25OE1km}; (2) generalizing the beam-bending effect to STWPs that are localized in both transverse spatial dimensions \cite{Yessenov22NC}; (3) testing the ability of bending STWPs to avoid line-of-sight targets; and (4) determining the maximum axial distance and transverse spatial displacement over which the STWP maintains its structure invariantly and its bending trajectory intact. We will present temporally resolved results for bending STWPs elsewhere. 

\textbf{Conclusion.} In summary, we have developed a new class of spatiotemporally structured self-accelerating optical fields in which the beam peak follows a curved trajectory. By appropriately modifying the spatiotemporal spectrum of an STWP, we arrange for the peak of the time-averaged intensity for a spatially \textit{symmetric} bending STWP to travel along a very general class of trajectories described by a power law. Uniquely, the acceleration rate and the power exponent are both independent of the spatial scale and shape of the beam profile. The flexibility available in the design of spatiotemporally structured optical fields suggests the possibility of producing beams capable of avoiding line-of-sight targets.

\section*{Funding}
U.S. Office of Naval Research (ONR) N00014-19-1-2192 and N00014-20-1-2789.

\section*{Data availability}
Data underlying the results presented in this paper are not publicly available at this time but may be obtained from the authors upon reasonable request.

\section*{Disclosures}
The authors declare no conflicts of interest.

\bibliography{diffraction}

\begin{thebibliography}{41}%
\makeatletter
\providecommand \@ifxundefined [1]{%
 \@ifx{#1\undefined}
}%
\providecommand \@ifnum [1]{%
 \ifnum #1\expandafter \@firstoftwo
 \else \expandafter \@secondoftwo
 \fi
}%
\providecommand \@ifx [1]{%
 \ifx #1\expandafter \@firstoftwo
 \else \expandafter \@secondoftwo
 \fi
}%
\providecommand \natexlab [1]{#1}%
\providecommand \enquote  [1]{``#1''}%
\providecommand \bibnamefont  [1]{#1}%
\providecommand \bibfnamefont [1]{#1}%
\providecommand \citenamefont [1]{#1}%
\providecommand \href@noop [0]{\@secondoftwo}%
\providecommand \href [0]{\begingroup \@sanitize@url \@href}%
\providecommand \@href[1]{\@@startlink{#1}\@@href}%
\providecommand \@@href[1]{\endgroup#1\@@endlink}%
\providecommand \@sanitize@url [0]{\catcode `\\12\catcode `\$12\catcode `\&12\catcode `\#12\catcode `\^12\catcode `\_12\catcode `\%12\relax}%
\providecommand \@@startlink[1]{}%
\providecommand \@@endlink[0]{}%
\providecommand \url  [0]{\begingroup\@sanitize@url \@url }%
\providecommand \@url [1]{\endgroup\@href {#1}{\urlprefix }}%
\providecommand \urlprefix  [0]{URL }%
\providecommand \Eprint [0]{\href }%
\providecommand \doibase [0]{https://doi.org/}%
\providecommand \selectlanguage [0]{\@gobble}%
\providecommand \bibinfo  [0]{\@secondoftwo}%
\providecommand \bibfield  [0]{\@secondoftwo}%
\providecommand \translation [1]{[#1]}%
\providecommand \BibitemOpen [0]{}%
\providecommand \bibitemStop [0]{}%
\providecommand \bibitemNoStop [0]{.\EOS\space}%
\providecommand \EOS [0]{\spacefactor3000\relax}%
\providecommand \BibitemShut  [1]{\csname bibitem#1\endcsname}%
\let\auto@bib@innerbib\@empty
\bibitem [{\citenamefont {Siviloglou}\ and\ \citenamefont {Christodoulides}(2007)}]{Siviloglou07OL}%
  \BibitemOpen
  \bibfield  {author} {\bibinfo {author} {\bibfnamefont {G.~A.}\ \bibnamefont {Siviloglou}}\ and\ \bibinfo {author} {\bibfnamefont {D.~N.}\ \bibnamefont {Christodoulides}},\ }\bibfield  {title} {\bibinfo {title} {Accelerating finite energy {A}iry beams},\ }\href@noop {} {\bibfield  {journal} {\bibinfo  {journal} {Opt. Lett.}\ }\textbf {\bibinfo {volume} {32}},\ \bibinfo {pages} {979} (\bibinfo {year} {2007})}\BibitemShut {NoStop}%
\bibitem [{\citenamefont {Siviloglou}\ \emph {et~al.}(2007)\citenamefont {Siviloglou}, \citenamefont {Broky}, \citenamefont {Dogariu},\ and\ \citenamefont {Christodoulides}}]{Siviloglou07PRL}%
  \BibitemOpen
  \bibfield  {author} {\bibinfo {author} {\bibfnamefont {G.~A.}\ \bibnamefont {Siviloglou}}, \bibinfo {author} {\bibfnamefont {J.}~\bibnamefont {Broky}}, \bibinfo {author} {\bibfnamefont {A.}~\bibnamefont {Dogariu}},\ and\ \bibinfo {author} {\bibfnamefont {D.~N.}\ \bibnamefont {Christodoulides}},\ }\bibfield  {title} {\bibinfo {title} {Observation of accelerating {A}iry beams},\ }\href@noop {} {\bibfield  {journal} {\bibinfo  {journal} {Phys. Rev. Lett.}\ }\textbf {\bibinfo {volume} {99}},\ \bibinfo {pages} {213901} (\bibinfo {year} {2007})}\BibitemShut {NoStop}%
\bibitem [{\citenamefont {Bandres}(2008)}]{Bandres08OL}%
  \BibitemOpen
  \bibfield  {author} {\bibinfo {author} {\bibfnamefont {M.}~\bibnamefont {Bandres}},\ }\bibfield  {title} {\bibinfo {title} {Accelerating beams},\ }\href@noop {} {\bibfield  {journal} {\bibinfo  {journal} {Opt. Lett.}\ }\textbf {\bibinfo {volume} {34}},\ \bibinfo {pages} {3791} (\bibinfo {year} {2008})}\BibitemShut {NoStop}%
\bibitem [{\citenamefont {Davis}\ \emph {et~al.}(2009)\citenamefont {Davis}, \citenamefont {Mitry}, \citenamefont {Bandres},\ and\ \citenamefont {Cottrell}}]{Davis09OE}%
  \BibitemOpen
  \bibfield  {author} {\bibinfo {author} {\bibfnamefont {J.~A.}\ \bibnamefont {Davis}}, \bibinfo {author} {\bibfnamefont {M.~J.}\ \bibnamefont {Mitry}}, \bibinfo {author} {\bibfnamefont {M.~A.}\ \bibnamefont {Bandres}},\ and\ \bibinfo {author} {\bibfnamefont {D.~M.}\ \bibnamefont {Cottrell}},\ }\bibfield  {title} {\bibinfo {title} {Observation of accelerating parabolic beams},\ }\href@noop {} {\bibfield  {journal} {\bibinfo  {journal} {Opt. Express}\ }\textbf {\bibinfo {volume} {16}},\ \bibinfo {pages} {12866} (\bibinfo {year} {2009})}\BibitemShut {NoStop}%
\bibitem [{\citenamefont {Efremidis}\ \emph {et~al.}(2019)\citenamefont {Efremidis}, \citenamefont {Chen}, \citenamefont {Segev},\ and\ \citenamefont {Christodoulides}}]{Efremidis19Optica}%
  \BibitemOpen
  \bibfield  {author} {\bibinfo {author} {\bibfnamefont {N.~K.}\ \bibnamefont {Efremidis}}, \bibinfo {author} {\bibfnamefont {Z.}~\bibnamefont {Chen}}, \bibinfo {author} {\bibfnamefont {M.}~\bibnamefont {Segev}},\ and\ \bibinfo {author} {\bibfnamefont {D.~N.}\ \bibnamefont {Christodoulides}},\ }\bibfield  {title} {\bibinfo {title} {Airy beams and accelerating waves: an overview of recent advances},\ }\href@noop {} {\bibfield  {journal} {\bibinfo  {journal} {Optica}\ }\textbf {\bibinfo {volume} {6}},\ \bibinfo {pages} {686} (\bibinfo {year} {2019})}\BibitemShut {NoStop}%
\bibitem [{\citenamefont {Polynkin}\ \emph {et~al.}(2009)\citenamefont {Polynkin}, \citenamefont {Kolesik}, \citenamefont {Moloney}, \citenamefont {Siviloglou},\ and\ \citenamefont {Christodoulides}}]{Polynkin09Science}%
  \BibitemOpen
  \bibfield  {author} {\bibinfo {author} {\bibfnamefont {P.}~\bibnamefont {Polynkin}}, \bibinfo {author} {\bibfnamefont {M.}~\bibnamefont {Kolesik}}, \bibinfo {author} {\bibfnamefont {J.~V.}\ \bibnamefont {Moloney}}, \bibinfo {author} {\bibfnamefont {G.~A.}\ \bibnamefont {Siviloglou}},\ and\ \bibinfo {author} {\bibfnamefont {D.~N.}\ \bibnamefont {Christodoulides}},\ }\bibfield  {title} {\bibinfo {title} {Curved plasma channel generation using ultra-intense {A}iry beams},\ }\href@noop {} {\bibfield  {journal} {\bibinfo  {journal} {Science}\ }\textbf {\bibinfo {volume} {324}},\ \bibinfo {pages} {229} (\bibinfo {year} {2009})}\BibitemShut {NoStop}%
\bibitem [{\citenamefont {Clerici}\ \emph {et~al.}(2015)\citenamefont {Clerici}, \citenamefont {Hu}, \citenamefont {Lassonde}, \citenamefont {Milián}, \citenamefont {Couairon}, \citenamefont {Christodoulides}, \citenamefont {Chen}, \citenamefont {Razzari}, \citenamefont {Vidal}, \citenamefont {L{\'e}gar{\'e}}, \citenamefont {Faccio},\ and\ \citenamefont {Morandotti}}]{Clerici15SciAdv}%
  \BibitemOpen
  \bibfield  {author} {\bibinfo {author} {\bibfnamefont {M.}~\bibnamefont {Clerici}}, \bibinfo {author} {\bibfnamefont {Y.}~\bibnamefont {Hu}}, \bibinfo {author} {\bibfnamefont {P.}~\bibnamefont {Lassonde}}, \bibinfo {author} {\bibfnamefont {C.}~\bibnamefont {Milián}}, \bibinfo {author} {\bibfnamefont {A.}~\bibnamefont {Couairon}}, \bibinfo {author} {\bibfnamefont {D.~N.}\ \bibnamefont {Christodoulides}}, \bibinfo {author} {\bibfnamefont {Z.}~\bibnamefont {Chen}}, \bibinfo {author} {\bibfnamefont {L.}~\bibnamefont {Razzari}}, \bibinfo {author} {\bibfnamefont {F.}~\bibnamefont {Vidal}}, \bibinfo {author} {\bibfnamefont {F.}~\bibnamefont {L{\'e}gar{\'e}}}, \bibinfo {author} {\bibfnamefont {D.}~\bibnamefont {Faccio}},\ and\ \bibinfo {author} {\bibfnamefont {R.}~\bibnamefont {Morandotti}},\ }\bibfield  {title} {\bibinfo {title} {Laser-assisted guiding of electric discharges around objects},\ }\href@noop {} {\bibfield  {journal} {\bibinfo  {journal} {Sci. Adv.}\ }\textbf {\bibinfo {volume} {1}},\ \bibinfo {pages}
  {e1400111} (\bibinfo {year} {2015})}\BibitemShut {NoStop}%
\bibitem [{\citenamefont {Vettenburg}\ \emph {et~al.}(2014)\citenamefont {Vettenburg}, \citenamefont {Dalgarno}, \citenamefont {Nylk}, \citenamefont {C.-Llad{\'o}}, \citenamefont {Ferrier}, \citenamefont {{\v C}i{\v z}m{\'a}r}, \citenamefont {Gunn-Moore},\ and\ \citenamefont {Dholakia}}]{Vettenburg2014NM}%
  \BibitemOpen
  \bibfield  {author} {\bibinfo {author} {\bibfnamefont {T.}~\bibnamefont {Vettenburg}}, \bibinfo {author} {\bibfnamefont {H.~I.~C.}\ \bibnamefont {Dalgarno}}, \bibinfo {author} {\bibfnamefont {J.}~\bibnamefont {Nylk}}, \bibinfo {author} {\bibfnamefont {C.}~\bibnamefont {C.-Llad{\'o}}}, \bibinfo {author} {\bibfnamefont {D.~E.~K.}\ \bibnamefont {Ferrier}}, \bibinfo {author} {\bibfnamefont {T.}~\bibnamefont {{\v C}i{\v z}m{\'a}r}}, \bibinfo {author} {\bibfnamefont {F.~J.}\ \bibnamefont {Gunn-Moore}},\ and\ \bibinfo {author} {\bibfnamefont {K.}~\bibnamefont {Dholakia}},\ }\bibfield  {title} {\bibinfo {title} {Light-sheet microscopy using an {A}iry beam},\ }\href@noop {} {\bibfield  {journal} {\bibinfo  {journal} {Nat. Meth.}\ }\textbf {\bibinfo {volume} {11}},\ \bibinfo {pages} {541} (\bibinfo {year} {2014})}\BibitemShut {NoStop}%
\bibitem [{\citenamefont {Subedi}\ \emph {et~al.}(2021)\citenamefont {Subedi}, \citenamefont {Yaraghi}, \citenamefont {Jung}, \citenamefont {Kukal}, \citenamefont {McDonald}, \citenamefont {Christodoulides},\ and\ \citenamefont {Vasdekis}}]{Subedi2021OE}%
  \BibitemOpen
  \bibfield  {author} {\bibinfo {author} {\bibfnamefont {N.~R.}\ \bibnamefont {Subedi}}, \bibinfo {author} {\bibfnamefont {S.}~\bibnamefont {Yaraghi}}, \bibinfo {author} {\bibfnamefont {P.~S.}\ \bibnamefont {Jung}}, \bibinfo {author} {\bibfnamefont {G.}~\bibnamefont {Kukal}}, \bibinfo {author} {\bibfnamefont {A.~G.}\ \bibnamefont {McDonald}}, \bibinfo {author} {\bibfnamefont {D.~N.}\ \bibnamefont {Christodoulides}},\ and\ \bibinfo {author} {\bibfnamefont {A.~E.}\ \bibnamefont {Vasdekis}},\ }\bibfield  {title} {\bibinfo {title} {Airy light-sheet {R}aman imaging},\ }\href@noop {} {\bibfield  {journal} {\bibinfo  {journal} {Opt. Express}\ }\textbf {\bibinfo {volume} {29}},\ \bibinfo {pages} {31941} (\bibinfo {year} {2021})}\BibitemShut {NoStop}%
\bibitem [{\citenamefont {Kaminer}\ \emph {et~al.}(2015)\citenamefont {Kaminer}, \citenamefont {Nemirovsky}, \citenamefont {Rechtsman}, \citenamefont {Bekenstein},\ and\ \citenamefont {Segev}}]{Kaminer15NPhys}%
  \BibitemOpen
  \bibfield  {author} {\bibinfo {author} {\bibfnamefont {I.}~\bibnamefont {Kaminer}}, \bibinfo {author} {\bibfnamefont {J.}~\bibnamefont {Nemirovsky}}, \bibinfo {author} {\bibfnamefont {M.}~\bibnamefont {Rechtsman}}, \bibinfo {author} {\bibfnamefont {R.}~\bibnamefont {Bekenstein}},\ and\ \bibinfo {author} {\bibfnamefont {M.}~\bibnamefont {Segev}},\ }\bibfield  {title} {\bibinfo {title} {Self-accelerating {D}irac particles and prolonging the lifetime of relativistic fermions},\ }\href@noop {} {\bibfield  {journal} {\bibinfo  {journal} {Nat. Phys.}\ }\textbf {\bibinfo {volume} {11}},\ \bibinfo {pages} {261} (\bibinfo {year} {2015})}\BibitemShut {NoStop}%
\bibitem [{\citenamefont {Kondakci}\ and\ \citenamefont {Abouraddy}(2017)}]{Kondakci17NP}%
  \BibitemOpen
  \bibfield  {author} {\bibinfo {author} {\bibfnamefont {H.~E.}\ \bibnamefont {Kondakci}}\ and\ \bibinfo {author} {\bibfnamefont {A.~F.}\ \bibnamefont {Abouraddy}},\ }\bibfield  {title} {\bibinfo {title} {Diffraction-free space-time light sheets},\ }\href@noop {} {\bibfield  {journal} {\bibinfo  {journal} {Nat. Photon.}\ }\textbf {\bibinfo {volume} {11}},\ \bibinfo {pages} {733} (\bibinfo {year} {2017})}\BibitemShut {NoStop}%
\bibitem [{\citenamefont {Yessenov}\ \emph {et~al.}(2022{\natexlab{a}})\citenamefont {Yessenov}, \citenamefont {Chen}, \citenamefont {Free}, \citenamefont {Johnson}, \citenamefont {Lavery}, \citenamefont {Alonso},\ and\ \citenamefont {Abouraddy}}]{Yessenov22NC}%
  \BibitemOpen
  \bibfield  {author} {\bibinfo {author} {\bibfnamefont {M.}~\bibnamefont {Yessenov}}, \bibinfo {author} {\bibfnamefont {Z.}~\bibnamefont {Chen}}, \bibinfo {author} {\bibfnamefont {J.}~\bibnamefont {Free}}, \bibinfo {author} {\bibfnamefont {E.~G.}\ \bibnamefont {Johnson}}, \bibinfo {author} {\bibfnamefont {M.~P.~J.}\ \bibnamefont {Lavery}}, \bibinfo {author} {\bibfnamefont {M.~A.}\ \bibnamefont {Alonso}},\ and\ \bibinfo {author} {\bibfnamefont {A.~F.}\ \bibnamefont {Abouraddy}},\ }\bibfield  {title} {\bibinfo {title} {Space-time wave packets localized in all dimensions},\ }\href@noop {} {\bibfield  {journal} {\bibinfo  {journal} {Nat. Commun.}\ }\textbf {\bibinfo {volume} {13}},\ \bibinfo {pages} {4573} (\bibinfo {year} {2022}{\natexlab{a}})}\BibitemShut {NoStop}%
\bibitem [{\citenamefont {Yessenov}\ \emph {et~al.}(2022{\natexlab{b}})\citenamefont {Yessenov}, \citenamefont {Hall}, \citenamefont {Schepler},\ and\ \citenamefont {Abouraddy}}]{Yessenov22AOP}%
  \BibitemOpen
  \bibfield  {author} {\bibinfo {author} {\bibfnamefont {M.}~\bibnamefont {Yessenov}}, \bibinfo {author} {\bibfnamefont {L.~A.}\ \bibnamefont {Hall}}, \bibinfo {author} {\bibfnamefont {K.~L.}\ \bibnamefont {Schepler}},\ and\ \bibinfo {author} {\bibfnamefont {A.~F.}\ \bibnamefont {Abouraddy}},\ }\bibfield  {title} {\bibinfo {title} {Space-time wave packets},\ }\href@noop {} {\bibfield  {journal} {\bibinfo  {journal} {Adv. Opt. Photon.}\ }\textbf {\bibinfo {volume} {14}},\ \bibinfo {pages} {455} (\bibinfo {year} {2022}{\natexlab{b}})}\BibitemShut {NoStop}%
\bibitem [{\citenamefont {Wong}\ and\ \citenamefont {Kaminer}(2017)}]{Wong17ACSP2}%
  \BibitemOpen
  \bibfield  {author} {\bibinfo {author} {\bibfnamefont {L.~J.}\ \bibnamefont {Wong}}\ and\ \bibinfo {author} {\bibfnamefont {I.}~\bibnamefont {Kaminer}},\ }\bibfield  {title} {\bibinfo {title} {Ultrashort tilted-pulsefront pulses and nonparaxial tilted-phase-front beams},\ }\href@noop {} {\bibfield  {journal} {\bibinfo  {journal} {ACS Photon.}\ }\textbf {\bibinfo {volume} {4}},\ \bibinfo {pages} {2257} (\bibinfo {year} {2017})}\BibitemShut {NoStop}%
\bibitem [{\citenamefont {Efremidis}(2017)}]{Efremidis17OL}%
  \BibitemOpen
  \bibfield  {author} {\bibinfo {author} {\bibfnamefont {N.~K.}\ \bibnamefont {Efremidis}},\ }\bibfield  {title} {\bibinfo {title} {Spatiotemporal diffraction-free pulsed beams in free-space of the {A}iry and {B}essel type},\ }\href@noop {} {\bibfield  {journal} {\bibinfo  {journal} {Opt. Lett.}\ }\textbf {\bibinfo {volume} {42}},\ \bibinfo {pages} {5038} (\bibinfo {year} {2017})}\BibitemShut {NoStop}%
\bibitem [{\citenamefont {Porras}(2001)}]{Porras01OL}%
  \BibitemOpen
  \bibfield  {author} {\bibinfo {author} {\bibfnamefont {M.~A.}\ \bibnamefont {Porras}},\ }\bibfield  {title} {\bibinfo {title} {Diffraction-free and dispersion-free pulsed beam propagation in dispersive media},\ }\href@noop {} {\bibfield  {journal} {\bibinfo  {journal} {Opt. Lett.}\ }\textbf {\bibinfo {volume} {26}},\ \bibinfo {pages} {1364} (\bibinfo {year} {2001})}\BibitemShut {NoStop}%
\bibitem [{\citenamefont {Kondakci}\ and\ \citenamefont {Abouraddy}(2019)}]{Kondakci19NC}%
  \BibitemOpen
  \bibfield  {author} {\bibinfo {author} {\bibfnamefont {H.~E.}\ \bibnamefont {Kondakci}}\ and\ \bibinfo {author} {\bibfnamefont {A.~F.}\ \bibnamefont {Abouraddy}},\ }\bibfield  {title} {\bibinfo {title} {Optical space-time wave packets of arbitrary group velocity in free space},\ }\href@noop {} {\bibfield  {journal} {\bibinfo  {journal} {Nat. Commun.}\ }\textbf {\bibinfo {volume} {10}},\ \bibinfo {pages} {929} (\bibinfo {year} {2019})}\BibitemShut {NoStop}%
\bibitem [{\citenamefont {Bhaduri}\ \emph {et~al.}(2020)\citenamefont {Bhaduri}, \citenamefont {Yessenov},\ and\ \citenamefont {Abouraddy}}]{Bhaduri20NP}%
  \BibitemOpen
  \bibfield  {author} {\bibinfo {author} {\bibfnamefont {B.}~\bibnamefont {Bhaduri}}, \bibinfo {author} {\bibfnamefont {M.}~\bibnamefont {Yessenov}},\ and\ \bibinfo {author} {\bibfnamefont {A.~F.}\ \bibnamefont {Abouraddy}},\ }\bibfield  {title} {\bibinfo {title} {Anomalous refraction of optical spacetime wave packets},\ }\href@noop {} {\bibfield  {journal} {\bibinfo  {journal} {Nat. Photon.}\ }\textbf {\bibinfo {volume} {14}},\ \bibinfo {pages} {416} (\bibinfo {year} {2020})}\BibitemShut {NoStop}%
\bibitem [{\citenamefont {Kondakci}\ and\ \citenamefont {Abouraddy}(2018{\natexlab{a}})}]{Kondakci18OL}%
  \BibitemOpen
  \bibfield  {author} {\bibinfo {author} {\bibfnamefont {H.~E.}\ \bibnamefont {Kondakci}}\ and\ \bibinfo {author} {\bibfnamefont {A.~F.}\ \bibnamefont {Abouraddy}},\ }\bibfield  {title} {\bibinfo {title} {Self-healing of space-time light sheets},\ }\href@noop {} {\bibfield  {journal} {\bibinfo  {journal} {Opt. Lett.}\ }\textbf {\bibinfo {volume} {43}},\ \bibinfo {pages} {3830} (\bibinfo {year} {2018}{\natexlab{a}})}\BibitemShut {NoStop}%
\bibitem [{\citenamefont {He}\ \emph {et~al.}(2022)\citenamefont {He}, \citenamefont {Guo},\ and\ \citenamefont {Xiao}}]{He22LPR}%
  \BibitemOpen
  \bibfield  {author} {\bibinfo {author} {\bibfnamefont {H.}~\bibnamefont {He}}, \bibinfo {author} {\bibfnamefont {C.}~\bibnamefont {Guo}},\ and\ \bibinfo {author} {\bibfnamefont {M.}~\bibnamefont {Xiao}},\ }\bibfield  {title} {\bibinfo {title} {Nondispersive space--time wave packets propagating in dispersive media},\ }\href@noop {} {\bibfield  {journal} {\bibinfo  {journal} {Laser Photon. Rev.}\ }\textbf {\bibinfo {volume} {16}},\ \bibinfo {pages} {2100634} (\bibinfo {year} {2022})}\BibitemShut {NoStop}%
\bibitem [{\citenamefont {Hall}\ and\ \citenamefont {Abouraddy}(2023{\natexlab{a}})}]{Hall23LPR}%
  \BibitemOpen
  \bibfield  {author} {\bibinfo {author} {\bibfnamefont {L.~A.}\ \bibnamefont {Hall}}\ and\ \bibinfo {author} {\bibfnamefont {A.~F.}\ \bibnamefont {Abouraddy}},\ }\bibfield  {title} {\bibinfo {title} {Canceling and inverting normal and anomalous group-velocity dispersion using space-time wave packets},\ }\href@noop {} {\bibfield  {journal} {\bibinfo  {journal} {Laser Photon. Rev.}\ }\textbf {\bibinfo {volume} {17}},\ \bibinfo {pages} {2200119} (\bibinfo {year} {2023}{\natexlab{a}})}\BibitemShut {NoStop}%
\bibitem [{\citenamefont {Clerici}\ \emph {et~al.}(2008)\citenamefont {Clerici}, \citenamefont {Faccio}, \citenamefont {Lotti}, \citenamefont {Rubino}, \citenamefont {Jedrkiewicz}, \citenamefont {Biegert},\ and\ \citenamefont {Trapani}}]{Clerici08OE}%
  \BibitemOpen
  \bibfield  {author} {\bibinfo {author} {\bibfnamefont {M.}~\bibnamefont {Clerici}}, \bibinfo {author} {\bibfnamefont {D.}~\bibnamefont {Faccio}}, \bibinfo {author} {\bibfnamefont {A.}~\bibnamefont {Lotti}}, \bibinfo {author} {\bibfnamefont {E.}~\bibnamefont {Rubino}}, \bibinfo {author} {\bibfnamefont {O.}~\bibnamefont {Jedrkiewicz}}, \bibinfo {author} {\bibfnamefont {J.}~\bibnamefont {Biegert}},\ and\ \bibinfo {author} {\bibfnamefont {P.~D.}\ \bibnamefont {Trapani}},\ }\bibfield  {title} {\bibinfo {title} {Finite-energy, accelerating {B}essel pulses},\ }\href@noop {} {\bibfield  {journal} {\bibinfo  {journal} {Opt. Express}\ }\textbf {\bibinfo {volume} {16}},\ \bibinfo {pages} {19807} (\bibinfo {year} {2008})}\BibitemShut {NoStop}%
\bibitem [{\citenamefont {Yessenov}\ and\ \citenamefont {Abouraddy}(2020)}]{Yessenov20PRLAccel}%
  \BibitemOpen
  \bibfield  {author} {\bibinfo {author} {\bibfnamefont {M.}~\bibnamefont {Yessenov}}\ and\ \bibinfo {author} {\bibfnamefont {A.~F.}\ \bibnamefont {Abouraddy}},\ }\bibfield  {title} {\bibinfo {title} {Accelerating and decelerating space-time wave packets in free space},\ }\href@noop {} {\bibfield  {journal} {\bibinfo  {journal} {Phys. Rev. Lett.}\ }\textbf {\bibinfo {volume} {125}},\ \bibinfo {pages} {233901} (\bibinfo {year} {2020})}\BibitemShut {NoStop}%
\bibitem [{\citenamefont {Li}\ and\ \citenamefont {Kawanaka}(2020)}]{Li20SR}%
  \BibitemOpen
  \bibfield  {author} {\bibinfo {author} {\bibfnamefont {Z.}~\bibnamefont {Li}}\ and\ \bibinfo {author} {\bibfnamefont {J.}~\bibnamefont {Kawanaka}},\ }\bibfield  {title} {\bibinfo {title} {Velocity and acceleration freely tunable straight-line propagation light bullet},\ }\href@noop {} {\bibfield  {journal} {\bibinfo  {journal} {Sci. Rep.}\ }\textbf {\bibinfo {volume} {10}},\ \bibinfo {pages} {11481} (\bibinfo {year} {2020})}\BibitemShut {NoStop}%
\bibitem [{\citenamefont {Hall}\ \emph {et~al.}(2022{\natexlab{a}})\citenamefont {Hall}, \citenamefont {Yessenov},\ and\ \citenamefont {Abouraddy}}]{Hall22OLAccel}%
  \BibitemOpen
  \bibfield  {author} {\bibinfo {author} {\bibfnamefont {L.~A.}\ \bibnamefont {Hall}}, \bibinfo {author} {\bibfnamefont {M.}~\bibnamefont {Yessenov}},\ and\ \bibinfo {author} {\bibfnamefont {A.~F.}\ \bibnamefont {Abouraddy}},\ }\bibfield  {title} {\bibinfo {title} {Arbitrarily accelerating space-time wave packets},\ }\href@noop {} {\bibfield  {journal} {\bibinfo  {journal} {Opt. Lett.}\ }\textbf {\bibinfo {volume} {47}},\ \bibinfo {pages} {694} (\bibinfo {year} {2022}{\natexlab{a}})}\BibitemShut {NoStop}%
\bibitem [{\citenamefont {{Allende M}otz}\ \emph {et~al.}(2020)\citenamefont {{Allende M}otz}, \citenamefont {Yessenov},\ and\ \citenamefont {Abouraddy}}]{Motz21PRA}%
  \BibitemOpen
  \bibfield  {author} {\bibinfo {author} {\bibfnamefont {A.~M.}\ \bibnamefont {{Allende M}otz}}, \bibinfo {author} {\bibfnamefont {M.}~\bibnamefont {Yessenov}},\ and\ \bibinfo {author} {\bibfnamefont {A.~F.}\ \bibnamefont {Abouraddy}},\ }\bibfield  {title} {\bibinfo {title} {Axial spectral encoding of space-time wave packets},\ }\href@noop {} {\bibfield  {journal} {\bibinfo  {journal} {Phys. Rev. Appl.}\ }\textbf {\bibinfo {volume} {15}},\ \bibinfo {pages} {024067} (\bibinfo {year} {2020})}\BibitemShut {NoStop}%
\bibitem [{\citenamefont {Hall}\ \emph {et~al.}(2025)\citenamefont {Hall}, \citenamefont {Yessenov}, \citenamefont {Romer},\ and\ \citenamefont {Abouraddy}}]{Hall25OL}%
  \BibitemOpen
  \bibfield  {author} {\bibinfo {author} {\bibfnamefont {L.~A.}\ \bibnamefont {Hall}}, \bibinfo {author} {\bibfnamefont {M.}~\bibnamefont {Yessenov}}, \bibinfo {author} {\bibfnamefont {M.~A.}\ \bibnamefont {Romer}},\ and\ \bibinfo {author} {\bibfnamefont {A.~F.}\ \bibnamefont {Abouraddy}},\ }\bibfield  {title} {\bibinfo {title} {Long-distance axial spectral encoding using space-time wave packets},\ }\href@noop {} {\bibfield  {journal} {\bibinfo  {journal} {Opt. Lett.}\ }\textbf {\bibinfo {volume} {50}},\ \bibinfo {pages} {in press} (\bibinfo {year} {2025})}\BibitemShut {NoStop}%
\bibitem [{\citenamefont {Liang}\ \emph {et~al.}(2023)\citenamefont {Liang}, \citenamefont {Liu}, \citenamefont {Luo}, \citenamefont {Chen},\ and\ \citenamefont {Deng}}]{Liang23OL}%
  \BibitemOpen
  \bibfield  {author} {\bibinfo {author} {\bibfnamefont {Z.}~\bibnamefont {Liang}}, \bibinfo {author} {\bibfnamefont {Y.}~\bibnamefont {Liu}}, \bibinfo {author} {\bibfnamefont {Y.}~\bibnamefont {Luo}}, \bibinfo {author} {\bibfnamefont {H.}~\bibnamefont {Chen}},\ and\ \bibinfo {author} {\bibfnamefont {D.}~\bibnamefont {Deng}},\ }\bibfield  {title} {\bibinfo {title} {Space-time wave packets with arbitrary transverse and longitudinal accelerations},\ }\href@noop {} {\bibfield  {journal} {\bibinfo  {journal} {Opt. Lett.}\ }\textbf {\bibinfo {volume} {48}},\ \bibinfo {pages} {2543} (\bibinfo {year} {2023})}\BibitemShut {NoStop}%
\bibitem [{\citenamefont {Hall}\ and\ \citenamefont {Abouraddy}(2024)}]{Hall24JOSAA}%
  \BibitemOpen
  \bibfield  {author} {\bibinfo {author} {\bibfnamefont {L.~A.}\ \bibnamefont {Hall}}\ and\ \bibinfo {author} {\bibfnamefont {A.~F.}\ \bibnamefont {Abouraddy}},\ }\bibfield  {title} {\bibinfo {title} {Universal angular-dispersion synthesizer},\ }\href@noop {} {\bibfield  {journal} {\bibinfo  {journal} {J. Opt. Soc. Am. A}\ }\textbf {\bibinfo {volume} {41}},\ \bibinfo {pages} {83} (\bibinfo {year} {2024})}\BibitemShut {NoStop}%
\bibitem [{\citenamefont {Romer}\ \emph {et~al.}(2025)\citenamefont {Romer}, \citenamefont {Hall},\ and\ \citenamefont {Abouraddy}}]{Romer25JO}%
  \BibitemOpen
  \bibfield  {author} {\bibinfo {author} {\bibfnamefont {M.~A.}\ \bibnamefont {Romer}}, \bibinfo {author} {\bibfnamefont {L.~A.}\ \bibnamefont {Hall}},\ and\ \bibinfo {author} {\bibfnamefont {A.~F.}\ \bibnamefont {Abouraddy}},\ }\bibfield  {title} {\bibinfo {title} {Synthesis and characterization of space-time light sheets: a tutorial},\ }\href@noop {} {\bibfield  {journal} {\bibinfo  {journal} {J. Opt.}\ }\textbf {\bibinfo {volume} {27}},\ \bibinfo {pages} {013501} (\bibinfo {year} {2025})}\BibitemShut {NoStop}%
\bibitem [{\citenamefont {Kondakci}\ and\ \citenamefont {Abouraddy}(2018{\natexlab{b}})}]{Kondakci18PRL}%
  \BibitemOpen
  \bibfield  {author} {\bibinfo {author} {\bibfnamefont {H.~E.}\ \bibnamefont {Kondakci}}\ and\ \bibinfo {author} {\bibfnamefont {A.~F.}\ \bibnamefont {Abouraddy}},\ }\bibfield  {title} {\bibinfo {title} {Airy wavepackets accelerating in space-time},\ }\href@noop {} {\bibfield  {journal} {\bibinfo  {journal} {Phys. Rev. Lett.}\ }\textbf {\bibinfo {volume} {120}},\ \bibinfo {pages} {163901} (\bibinfo {year} {2018}{\natexlab{b}})}\BibitemShut {NoStop}%
\bibitem [{\citenamefont {Wong}(2021)}]{Wong21OE}%
  \BibitemOpen
  \bibfield  {author} {\bibinfo {author} {\bibfnamefont {L.~J.}\ \bibnamefont {Wong}},\ }\bibfield  {title} {\bibinfo {title} {Propagation-invariant space-time caustics of light},\ }\href@noop {} {\bibfield  {journal} {\bibinfo  {journal} {Opt. Express}\ }\textbf {\bibinfo {volume} {29}},\ \bibinfo {pages} {30682} (\bibinfo {year} {2021})}\BibitemShut {NoStop}%
\bibitem [{\citenamefont {Yessenov}\ \emph {et~al.}(2019)\citenamefont {Yessenov}, \citenamefont {Bhaduri}, \citenamefont {Kondakci}, \citenamefont {Meem}, \citenamefont {Menon},\ and\ \citenamefont {Abouraddy}}]{Yessenov19Optica}%
  \BibitemOpen
  \bibfield  {author} {\bibinfo {author} {\bibfnamefont {M.}~\bibnamefont {Yessenov}}, \bibinfo {author} {\bibfnamefont {B.}~\bibnamefont {Bhaduri}}, \bibinfo {author} {\bibfnamefont {H.~E.}\ \bibnamefont {Kondakci}}, \bibinfo {author} {\bibfnamefont {M.}~\bibnamefont {Meem}}, \bibinfo {author} {\bibfnamefont {R.}~\bibnamefont {Menon}},\ and\ \bibinfo {author} {\bibfnamefont {A.~F.}\ \bibnamefont {Abouraddy}},\ }\bibfield  {title} {\bibinfo {title} {Non-diffracting broadband incoherent space-time fields},\ }\href@noop {} {\bibfield  {journal} {\bibinfo  {journal} {Optica}\ }\textbf {\bibinfo {volume} {6}},\ \bibinfo {pages} {598} (\bibinfo {year} {2019})}\BibitemShut {NoStop}%
\bibitem [{\citenamefont {Yessenov}\ and\ \citenamefont {Abouraddy}(2019)}]{Yessenov19OL}%
  \BibitemOpen
  \bibfield  {author} {\bibinfo {author} {\bibfnamefont {M.}~\bibnamefont {Yessenov}}\ and\ \bibinfo {author} {\bibfnamefont {A.~F.}\ \bibnamefont {Abouraddy}},\ }\bibfield  {title} {\bibinfo {title} {Changing the speed of optical coherence in free space},\ }\href@noop {} {\bibfield  {journal} {\bibinfo  {journal} {Opt. Lett.}\ }\textbf {\bibinfo {volume} {44}},\ \bibinfo {pages} {5125} (\bibinfo {year} {2019})}\BibitemShut {NoStop}%
\bibitem [{\citenamefont {Saari}\ and\ \citenamefont {Reivelt}(2004)}]{Saari04PRE}%
  \BibitemOpen
  \bibfield  {author} {\bibinfo {author} {\bibfnamefont {P.}~\bibnamefont {Saari}}\ and\ \bibinfo {author} {\bibfnamefont {K.}~\bibnamefont {Reivelt}},\ }\bibfield  {title} {\bibinfo {title} {Generation and classification of localized waves by {L}orentz transformations in {F}ourier space},\ }\href@noop {} {\bibfield  {journal} {\bibinfo  {journal} {Phys. Rev. E}\ }\textbf {\bibinfo {volume} {69}},\ \bibinfo {pages} {036612} (\bibinfo {year} {2004})}\BibitemShut {NoStop}%
\bibitem [{\citenamefont {Longhi}(2004)}]{Longhi04OE}%
  \BibitemOpen
  \bibfield  {author} {\bibinfo {author} {\bibfnamefont {S.}~\bibnamefont {Longhi}},\ }\bibfield  {title} {\bibinfo {title} {Gaussian pulsed beams with arbitrary speed},\ }\href@noop {} {\bibfield  {journal} {\bibinfo  {journal} {Opt. Express}\ }\textbf {\bibinfo {volume} {12}},\ \bibinfo {pages} {935} (\bibinfo {year} {2004})}\BibitemShut {NoStop}%
\bibitem [{\citenamefont {Yessenov}\ and\ \citenamefont {Abouraddy}(2023)}]{Yessenov23PRA}%
  \BibitemOpen
  \bibfield  {author} {\bibinfo {author} {\bibfnamefont {M.}~\bibnamefont {Yessenov}}\ and\ \bibinfo {author} {\bibfnamefont {A.~F.}\ \bibnamefont {Abouraddy}},\ }\bibfield  {title} {\bibinfo {title} {Relativistic transformations of quasi-monochromatic optical beams},\ }\href@noop {} {\bibfield  {journal} {\bibinfo  {journal} {Phys. Rev. A}\ }\textbf {\bibinfo {volume} {107}},\ \bibinfo {pages} {042221} (\bibinfo {year} {2023})}\BibitemShut {NoStop}%
\bibitem [{\citenamefont {Ramsey}\ \emph {et~al.}(2023)\citenamefont {Ramsey}, \citenamefont {{Di P}iazza}, \citenamefont {Formanek}, \citenamefont {Franke}, \citenamefont {Froula}, \citenamefont {Malaca}, \citenamefont {Mori}, \citenamefont {Pierce}, \citenamefont {Simpson}, \citenamefont {Vieira}, \citenamefont {Vranic}, \citenamefont {Weichman},\ and\ \citenamefont {Palastro}}]{Ramsey23PRA}%
  \BibitemOpen
  \bibfield  {author} {\bibinfo {author} {\bibfnamefont {D.}~\bibnamefont {Ramsey}}, \bibinfo {author} {\bibfnamefont {A.}~\bibnamefont {{Di P}iazza}}, \bibinfo {author} {\bibfnamefont {M.}~\bibnamefont {Formanek}}, \bibinfo {author} {\bibfnamefont {P.}~\bibnamefont {Franke}}, \bibinfo {author} {\bibfnamefont {D.~H.}\ \bibnamefont {Froula}}, \bibinfo {author} {\bibfnamefont {B.}~\bibnamefont {Malaca}}, \bibinfo {author} {\bibfnamefont {W.~B.}\ \bibnamefont {Mori}}, \bibinfo {author} {\bibfnamefont {J.~R.}\ \bibnamefont {Pierce}}, \bibinfo {author} {\bibfnamefont {T.~T.}\ \bibnamefont {Simpson}}, \bibinfo {author} {\bibfnamefont {J.}~\bibnamefont {Vieira}}, \bibinfo {author} {\bibfnamefont {M.}~\bibnamefont {Vranic}}, \bibinfo {author} {\bibfnamefont {K.}~\bibnamefont {Weichman}},\ and\ \bibinfo {author} {\bibfnamefont {J.~P.}\ \bibnamefont {Palastro}},\ }\bibfield  {title} {\bibinfo {title} {Exact solutions for the electromagnetic fields of a flying focus},\ }\href@noop {} {\bibfield  {journal} {\bibinfo
  {journal} {Phys. Rev. A}\ }\textbf {\bibinfo {volume} {107}},\ \bibinfo {pages} {013513} (\bibinfo {year} {2023})}\BibitemShut {NoStop}%
\bibitem [{\citenamefont {Hall}\ and\ \citenamefont {Abouraddy}(2023{\natexlab{b}})}]{Hall23NP}%
  \BibitemOpen
  \bibfield  {author} {\bibinfo {author} {\bibfnamefont {L.~A.}\ \bibnamefont {Hall}}\ and\ \bibinfo {author} {\bibfnamefont {A.~F.}\ \bibnamefont {Abouraddy}},\ }\bibfield  {title} {\bibinfo {title} {Observation of optical de {B}roglie-{M}ackinnon wave packets},\ }\href@noop {} {\bibfield  {journal} {\bibinfo  {journal} {Nat. Phys.}\ }\textbf {\bibinfo {volume} {19}},\ \bibinfo {pages} {435} (\bibinfo {year} {2023}{\natexlab{b}})}\BibitemShut {NoStop}%
\bibitem [{\citenamefont {Yessenov}\ \emph {et~al.}(2024)\citenamefont {Yessenov}, \citenamefont {Romer}, \citenamefont {Ichiji},\ and\ \citenamefont {Abouraddy}}]{Yessenov24PRA}%
  \BibitemOpen
  \bibfield  {author} {\bibinfo {author} {\bibfnamefont {M.}~\bibnamefont {Yessenov}}, \bibinfo {author} {\bibfnamefont {M.}~\bibnamefont {Romer}}, \bibinfo {author} {\bibfnamefont {N.}~\bibnamefont {Ichiji}},\ and\ \bibinfo {author} {\bibfnamefont {A.~F.}\ \bibnamefont {Abouraddy}},\ }\bibfield  {title} {\bibinfo {title} {Experimental realization of lorentz boosts of space-time wave packets},\ }\href@noop {} {\bibfield  {journal} {\bibinfo  {journal} {Phys. Rev. A}\ }\textbf {\bibinfo {volume} {109}},\ \bibinfo {pages} {013509} (\bibinfo {year} {2024})}\BibitemShut {NoStop}%
\bibitem [{\citenamefont {Hall}\ \emph {et~al.}(2022{\natexlab{b}})\citenamefont {Hall}, \citenamefont {Romer}, \citenamefont {Turo}, \citenamefont {Hayward}, \citenamefont {Menon},\ and\ \citenamefont {Abouraddy}}]{Hall25OE1km}%
  \BibitemOpen
  \bibfield  {author} {\bibinfo {author} {\bibfnamefont {L.~A.}\ \bibnamefont {Hall}}, \bibinfo {author} {\bibfnamefont {M.~A.}\ \bibnamefont {Romer}}, \bibinfo {author} {\bibfnamefont {B.~L.}\ \bibnamefont {Turo}}, \bibinfo {author} {\bibfnamefont {T.~M.}\ \bibnamefont {Hayward}}, \bibinfo {author} {\bibfnamefont {R.}~\bibnamefont {Menon}},\ and\ \bibinfo {author} {\bibfnamefont {A.~F.}\ \bibnamefont {Abouraddy}},\ }\bibfield  {title} {\bibinfo {title} {Space-time wave packets propagating a kilometer in air},\ }\href@noop {} {\bibfield  {journal} {\bibinfo  {journal} {arXiv:2209.03309}\ } (\bibinfo {year} {2022}{\natexlab{b}})}\BibitemShut {NoStop}%
\end{thebibliography}%

\end{document}